\documentclass{article}
\usepackage[utf8]{inputenc}
\usepackage[section]{placeins}
\usepackage[export]{adjustbox}
\usepackage[parfill]{parskip}
\usepackage{graphicx}
\usepackage{amsmath}
\usepackage{amsfonts} 
\usepackage{hyperref}  
\usepackage{authblk}
\usepackage{cleveref}
\hypersetup{pdfborder={0 0 0} } 
\usepackage[a4paper, total={6in, 8in}]{geometry}
\graphicspath{ {./images/} }

\usepackage[
backend=biber,
style=authoryear-comp,
]{biblatex}

\title{An agent-based model of the 2020 international policy diffusion in response to the COVID-19 pandemic with particle filter}

\author[1]{Yannick Oswald\thanks{corresponding author: y-oswald@web.de}}
\author[1]{Keiran Suchak}
\author[1]{Nick Malleson}
\affil{School of Geography, University of Leeds}

\date{January 2023}

\begin{document}

\maketitle

\begin{abstract}
Global problems, such as pandemics and climate change, require rapid international coordination and diffusion of policy.
These phenomena are rare however, with one notable example being the international policy response to the COVID-19 pandemic in early 2020.
Here we build an agent-based model of this rapid policy diffusion, where countries constitute the agents and with the principal mechanism for diffusion being peer mimicry.
Since it is challenging to predict accurately the policy diffusion curve, we utilize data assimilation, that is an ``on-line'' feed of data to constrain the model against observations.
The specific data assimilation algorithm we apply is a particle filter because of its convenient implementation, its ability to handle categorical variables and because the model is not overly computationally expensive, hence a more efficient algorithm is not required.
We find that the model alone is able to predict the policy diffusion relatively well with an ensemble of at least 100 simulation runs.
The particle filter however improves the fit to the data, reliably so from 500 runs upwards, and increasing filtering frequency results in improved prediction.
\end{abstract}

\section{Introduction}

Several global challenges hinge on an international coordination of policy. Climate change requires a global fossil fuel phase out \parencite{Shukla2022}. Prevention of the next pandemic requires a cohesive bio-security strategy as well as minimum interference with ecosystems \parencite{Morse2012}. International security requires negotiation and a willingness to resolve conflict by peaceful means instead of resorting to violence \parencite{Bercovitch2009}. Moreover, many of these challenges are immensely time pressing. With respect to climate change, for instance, only few years remain to achieve international climate targets.

There has been one recent, yet already historic, instance of international policy coordination that was surprisingly rapid, decisive and homogeneous - the implementation of full-scale lockdowns as a response to the COVID-19 pandemic. There was an initial period of action delay, and in some cases even denial of the crisis, from roughly December 2019 to February 2020, but in March 2020 nearly every country in the world implemented stringent lockdowns, including closures of public venues and schools, mask wearing duties and mandatory home office. On the first of March 2020 only around 8 \% of countries had implemented such stringent measures, by the end of March 2020 more than 90\% had \parencite{owidcoronavirus}.

The pandemic has been scientifically scrutinized from many disciplinary viewpoints and data collection expanded almost universally across fields related to the pandemic. International and freely accessible virus case trackers emerged as well as most capable national governments swiftly implemented regional and sub-regional case tracking. Moreover, prompted by the drastic and widely perceived shift in the political landscape, various policy response trackers originated, some focusing only on the stringency of direct virus control mechanisms,  others also on the wider fiscal policy landscape as well as state-control mechanisms monitoring political variables such as democratic freedom \parencite{Daly2020}. There are large amounts of research into the biology of the virus and the epidemiology of COVID-19 and from a social scientific viewpoint, there is plenty of research on the economic effects of pandemic policies \parencite{Chetty2020}. 

The political science of the lockdown policies has comparatively received less attention and the rapid tipping point in March has been debated little from computational and simulation-oriented viewpoints. There is some quantitative work on the drivers of lockdown policies. To name one notable study, \textcite{Sebhatu2020} emphasized the remarkable degree of homogeneity in lockdown adoption by largely heterogeneous countries in a short amount of time. They operationalized the degree of lockdown employing five metrics, namely ``information campaigns'', ``school closures'', ``workplace closures'', ``cancellation of public events'' and ``restrictions on internal mobility''. They found that there are some country specific drivers, such as the population density and the degree of authoritarianism, that fostered the early adoption of lockdown policies. Democracies were generally slower to adopt stringent lockdowns. Their main result is that none of these drivers, nor the actual COVID case numbers in a nation are sufficient explanations for the rapid lockdown adoption in March 2020. Instead, they conclude that this diffusion of policy resembles peer-adoption processes in which countries mimicked other countries to cope with the threat and uncertainty posed by the pandemic. Democracies stand out again because, while slow to come up with their own lockdown initiative, they exhibited higher sensitivity to adoption through mimicking peers.

This insight suggests an agent-based modelling (ABM) approach to further elucidate the diffusion mechanisms and perhaps generalize them. Innovation and cultural norm diffusion models are a widespread application of ABMs and continue to be a very active field of ABM research \parencite{kiesling2012agent}. These models usually employ some kind of social network structure on which the diffusion happens and achieve respectable approximations to real-world diffusion processes \parencite{bohlmann2010effects, xiao2016forecasting}. Diffusion-oriented ABMs also have been extensively applied to the spread of the COVID-19 virus itself and the effect of policy interventions on the spread \parencite{lorig2021agent, kai2020universal,kerr2021covasim, cuevas2020agent}.


Despite this volume of existing research, there have been no agent-based studies of the international processes that underlie the formation of lockdown policies, which is a notable gap in the literature. Two reasons for this might be that: (i) modelling the complex social networks and dynamics within diverse governmental institutions that ultimately lead to the adoption of lockdown policies would be extremely ambitious at an international scale; and (ii) there are little to no data available that could reveal how governmental decision making is actually processed. However, if one of the most important drivers of lockdown policy adoption is peer mimicry, then a model of international lockdown policy may not need to consider the vast complexity and variety of national decision-making structures. Hence we contribute to the literature on lockdown policy diffusion by developing the first agent-based model of the adoption of lockdown policies, with agents representing countries. We focus on March 2020 and initialize the model based on real-world data from 1$^{st}$ March. The aims of our work are to (i) better understand the influence of peer mimicry on national lockdown adoption and (ii) build a model that could be used in the future to predict short-term national policy changes before they occur.

With respect to the second aim, an additional challenge is how to capture the rapidity with which countries adopted lockdown-related policies. There was a tipping point over two weeks in March 2020 when the number of countries adopting the most extreme policies went from around 20 to more than 160 (discussed in detail in Section~\ref{sec:policy_data}). It is highly unlikely that any model will be sufficiently accurate to capture the precise days when countries changed their policies. But due to the rapidity at which countries adopted lockdowns, even small model prediction errors will lead to the simulated tipping point occurring too early, or too late, and subsequently very poor model performance. Hence we employ a novel methodological feature with respect to ABMs to try to align the model with the evolution of the real system; that of data-assimilation (DA). This means we constrain the model by regularly updated observations covering the state of the real-world, thereby improving the accuracy of future model predictions. In our case, these new observations include whether a country has adopted a lockdown or not. The DA approach we choose is a particle filter since it is well-suited for highly non-linear systems and is able to cope with categorical variables; a feature our model relies on. Moreover it has been applied to a few ABMs already, although they were mostly simulating pedestrian dynamics \parencite{wang_data_2015, malleson2020, ternes2022data}. Data-assimilation with a diffusion-oriented ABM let alone a policy focus is a novel application. Our goal here is to test this new application, highlight advantages and point to challenges. We find that the particle filter generally improves the prediction, through reducing the mean-squared error (MSE) of predictions with the real-world data as a reference, and also increases the reliability of the forecast by reducing the variance of estimates.

The remainder of the paper is structured as follows: We review the relevant literature, then introduce the model, outline methods and data applied, show experimental results and discuss the findings.

\section{Background}

\subsection{Policy diffusion across nations}\label{sec:policy_diffusion}

Our model is essentially one of policy diffusion across nations -- a  major research topic in political science. International policy diffusion is a complex field and the exact mechanisms for diffusion are debated and also vary from policy issue to policy issue. Mechanisms in consideration include ideological similarity between countries, strategic and economic competition, learning and coercion \parencite{baybeck2011strategic, gilardi2016four, dobbin2007global}. Generically, international policy diffusion may be be defined as ``...when government policy decisions in a given country are systematically conditioned by prior policy choices made in other countries...'' \parencite{simmons_dobbin_garrett_2006}. Given this definition, the Covid-lockdown adoption across the world arguably constitutes such a phenomenon. A substantial amount of research has also considered sub-national policy diffusion, prominently across states in the United States \parencite{shipan2008mechanisms} but recently for instance also across provinces in China \parencite{zhang2019multiple}. Themes that have been investigated under the umbrella of policy diffusion include environmental policies and technologies \parencite{zimm2021improving}, social security reform \parencite{brooks2004role}, health policies \parencite{gautier2021transnational}, mental health policies \parencite{shen2014cross}, the spread of democracy \parencite{o1998diffusion}, economic policies and economic liberalization \parencite{simmons2004globalization} and more.

Policy diffusion may unfold over widely varying timescales, yet most of the time it has occured slowly on the order of decades or even centuries. For example the world took 150 years to go from ~5\% of countries being democratic to 60\% being democratic \parencite{owiddemocracy}. Admittedly, the transition to democracy is not a single policy but rather a deep and fundamental change in a countries political fabric, but even single normative and legal reforms can take decades to diffuse. For example, \textcite{tews2003diffusion} show that soil, air and water protection laws all took four to five decades to diffuse across OECD-countries as well as Central and Eastern European countries.

Importantly for our purposes, some policy diffusion research has specifically focused on the COVID-19 emergency in early 2020 and the observed \textit{rapid} policy diffusion. 
\textcite{lundgren2020emergency} for example have investigated the diffusion of ``state of emergency (SOE) declarations''. They show that the declarations in early 2020 follow a typical diffusion curve but maxed out at roughly 50\% of all countries world-wide. They also ascertain the drivers of SOE declarations. According to their results regional clustering occurs. Countries opted for SOE, if neighboring countries did as well and the probability to adopt SOE depends on the degree of democracy and pandemic preparedness.  

As aforementioned, \textcite{Sebhatu2020} found that there are some internal drivers of lockdown policies (e.g. population density and level of democracy) but none of those alone suffices to explain the archetypal diffusion curve observed in March 2020. Instead they argue that peer mimicry must have driven the process. This hypothesis is corroborated by other studies. For example \textcite{mistur2020contagious} employ fixed effect models on panel data throughout 2020 and demonstrate that mimicry of geographical neighbors and political peers in addition to having a language in common are the principal drivers for countries to introduce, or abandon again, social distancing measures.

There are also a small number of agent-based models applied to generic issues of international policy diffusion. For example, \textcite{rapaport2009puzzle} were motivated by the diffusion of central bank autonomy around the world. They found that between 1990 and 2008, 84 countries strongly increased the independence of their central banks; approximately 5 countries per year. This is a steep rate of change compared to the status quo before that, yet far from the rate of policy change observed in 2020 during the pandemic. Instead of just focusing on this particular policy diffusion phenomenon, they attempt to build a generic agent-based model of national policy diffusion. Their model is informed by Kingdon's theory of national policy formation who considered policy issues and policy solutions to have a life of their own and arguably perceived them undergoing an evolutionary process (according to Kingdon, policy solutions have their own fitness score for instance) \parencite{kingdon1984agendas}. Crucially, \textcite{rapaport2009puzzle} integrated these politically endogenous mechanisms together with external mechanisms. External mechanisms refer to when agents, the countries, look to other countries in their ``zone of influence''. Here ``zone of influence'' is an umbrella term for geographical and ideological proximity. Eventually a mixture of internal and external factors determines whether a country adopts a new policy or not, represented as a binary variable (`yes' or `no') in their model. Moreover, there is the agent-based framework by \textcite{luyet2011policy}. They build a model inspired by earlier attempts to formalize national diffusion theory (mostly by \textcite{braun2006taking}). In this work the probability that a country adopts a policy is influenced by the ``effectiveness'' of the policy in other regions, domestic institutional constraints and the proportion of geographic neighbors that already have adopted the policy. The study then proceeds to test the impact of parameters on the S-diffusion curve as an outcome but stops short of relating to any empirical case. 

The only other agent-based approach to national policy diffusion that we are aware of is by \textcite{ring2014agent}. Ring argues that diffusion of policies can happen through three structural properties of the international political landscape: hierarchy, neighborhood and identity. Neighborhood refers to geographical proximity and identity to ideological proximity, which is similar to other models. Hierarchy is one feature that sets his model apart. Ring argues that few countries are high ranking in the international order and most countries low ranking. He then further maps these properties on the four mechanisms discussed in the diffusion-literature (coercion, learning, emulation and competition). He concludes that the four mechanisms vary based on the speed of diffusion and whether they actually produce an archetypal S-curve. According to Ring, learning and competition do, while coercion and, surprisingly, emulation produce rather flat curves. 

\subsection{Social networks and agent-based diffusion models}

While not a dominant method in political science overall, a relatively popular application of ABMs and computational approaches in political studies is the explanation of polarization or coalition formation on social networks; compare for instance \textcite{li2017agent, leifeld2014polarization, batista2019migrant}. Since we build a model in order to explain a diffusion process, it is worth reviewing (very) briefly what the main insights are from this line of literature. Social network diffusion models originated essentially from even broader network research. This research has initially been driven by the physics community who noted that if node degree in networks is made to follow certain probability distributions then interesting networks can emerge. A popular example is a scale-free network whose degree distribution follows a power law \parencite{barabasi1999emergence}. The generative mechanisms for these degree distributions are also a large research field in themselves but of lesser concern to us here. Eventually, varied types of networks have been applied to diverse kinds of social diffusion processes including political opinions and technology. For example it has been shown that centrally positioned individuals in social networks can significantly influence the polarization of opinions and that social media shapes networks in a way that enables diffusion \parencite{kandiah2012pagerank, lu2021big}.  
A broad result from this line of research is that the social network structure matters \parencite{li2017agent}. Networks with some kind of central nodes and heterogeneous degree distribution enable faster and more stable diffusion patterns than uniformly arranged networks. Local clustering for instance is one network structure that enables rapid diffusion \parencite{kreindler2014rapid}. Comparing this insight to the international policy diffusion literature, there might be similar effects between countries. Local groups of countries and hierarchies among countries likely influence speed and nature of policy diffusion. 

\subsection{Data assimilation for agent-based models}

Agent-based models (ABMs) are now an established tool to model complex systems. ABMs have succeeded in illuminating system dynamics across many disciplines from chemistry and biology to economics, geography and sustainability sciences \parencite{axtell2022agent}. However, as with any model, ABMs have disadvantages. For example, it remains difficult to produce accurate forecasts because even if a model describes the core dynamics of some system in the past, it might not capture the gradually evolving mechanics over time. Any complex system, particularly complex social systems, are subject to tiny disturbances from a large variety of sources that can result in great differences compared to past behaviour -- this is the essence of chaos theory.

With big data emerging as an ubiquitous feature of our time, computational scientists have a tool at hand to constrain their models, and ultimately to control the chaos (at least to a certain degree). This is where data assimilation comes in. Data assimilation is a set of algorithms originally developed in weather forecasting in which real-time observations are integrated into a model on a continuous basis \parencite{kalnay2003atmospheric}. These real-time observations constrain the model against the evolving true state of the system \parencite{malleson2020}. 
Examples of the use of agent-based models with data assimilation are extremely rare. Only relatively recent literature applies particle filters \parencite{wang_data_2015, ternes2022data, malleson2020, hu2022data, lueck_who_2019}, other sequential Monte-Carlo sampling techniques \parencite{tang2022data} and varieties of the Kalman Filter \parencite{CLAY2021102386, ward2016dynamic} to ABMs, typically for crowd simulation or more general population movement. 

It is challenging to combine ABMs with data assimilation because the character of the model is distinct from a single equation or a system of equations. In ABMs, agents might possess a diverse set of behavioural parameters including numeric and categorical ones. Data assimilation algorithms however have been developed to optimise continuous numerical variables in weather and climate modelling \parencite{ternes2022data}. ABMs are essentially discrete event simulations \parencite{gatti2018agent}. Entities in the model behave according to rules. The rules are often stochastic and also determine an agents behaviour as a result of interaction with other agents. Hence, ABMs often become more complex at scale, particularly so if scaling up the number of agents implies scaling up the number of interactions. For example, \textcite{malleson2020} found, in their model of pedestrians, that 
data assimilation with a particle filter became extremely computationally costly because, as the number of agents in the environment increases, exponentially more particles are required to achieve a constant (low) error rate. 
In addition, \textcite{ternes2022data} found difficulties using a particle filter in an agent-based crowd simulation because \textit{particle deprivation} meant that the algorithm was not able to adequately search the space of possible model trajectories and ruled out models that would ultimately have predicted the system well. \textcite{ternes2022data} proposed a resolution to this problem by filtering some parts of the model state space but not others.
Variations of the Kalman Filter are potentially more efficient than the particle filter, but cannot estimate categorical parameters, although \textcite{CLAY2021102386} propose a `reversible jump' mechanism to get round this drawback.
Overall, for this work the particle filter appears to be the most suitable method as the proposed model of international lockdown behaviour only includes a fixed number of 164 agents (countries) and is much quicker to execute than a complex crowd model. Hence it should be possible to include a sufficiently large number of particles to avoid the problems identified by \textcite{ternes2022data} without becoming overly computationally expensive and requiring a more efficient algorithm such as \textcite{CLAY2021102386}.

\section{Model and methods}

\subsection{Model description}

The model is a data-driven agent-based model implemented in Python-MESA. The principal idea of the model is that the diffusion of lockdown policy across countries can be described independently of the actual COVID-case numbers across countries, at least for the period of interest which is March 2020. The main diffusion mechanism is that countries take note of which other countries already have adopted a lockdown and, if those include countries sufficiently similar to oneself, then they are likely to adopt a lockdown themselves. In other words, countries mimic countries that are similar. The model makes a simplification in that agents (countries) can either have adopted a lockdown or not, so the dependent variable of interest is binary. In reality, of course, there were varying degrees of lockdown intensity and stringency. However we observe in spring 2020 that only two outcomes really matter matter: `no lockdown' or `complete lockdown'. This is described in detail in Section~\ref{sec:policy_data}.
There are 164 agents in the model since for 164 countries we were able to collect sufficiently comprehensive data on the COVID policy response as well as country specific variables such as national income and degree of democracy.

Similarity between countries is measured along three dimensions: national income, degree of democracy and geographical location. National income is captured through Gross Domestic Product per capita in Purchasing Power Parity \parencite{worldbankGDP}, degree of democracy through the Democracy Index by the Economist Intelligence Unit \parencite{economistDemocracy2019} and geographical location simply through latitude and longitude of a country's capital \parencite{listCountriesCapitals}. 

For measuring similarity, we use an equally-weighted average of each of the the three dimensions. In each case, the quantities pertaining to a dimension are normalised on the unit interval $[0, 1]$.
We consider this similarity measure as a distance between two countries.
The lower the distance between two countries, the more similar thye are.
In formal terms, the distance, $d_{ij}$, between country $i$ and country $j$ is:
\begin{align}\label{first_equation}
    d_{ij} = \frac{1}{3}
             \left(
                \underbrace{\frac{(x_i - x_j)}{(x_{max} - x_{min})}}_{\text{income similarity}}
                +
                \underbrace{\frac{(y_i - y_j)}{(y_{max} - y_{min})}}_{\text{political similarity}}
                +
                \underbrace{\frac{H(z_i,z_j)}{H_{max}}}_{\text{geo. proximity}}
             \right)
\end{align}
where $x_i$ is the national income of country $i$, $y_i$ the degree of democracy of country $i$, $z_i$ is the location (in terms of latitude and longitude) of the capital of country $i$, and $H(a, b)$ denotes the haversine-formula for the distance between two points, $a$ and $b$.
Time indices are omitted for clarity in \cref{first_equation} as the distance between countries remains constant over time.
A variable over time is the number of countries who already adopted a lockdown, so at every time step this distance is evaluated for a new set of countries by each agent. 



At any given time-step, a country can be in one of two lockdown statuses: in ``lockdown'' or ``not in lockdown''.
We denote the binary lockdown status of the agent representing country $i$ at time $t$ as $\theta_t (i)$:
\begin{equation}\label{second_equation}
\theta_t(i)=
    \begin{cases}
        1 & \text{if } \text{country $i$ is in lockdown },\\
        0 & \text{if } \text{country $i$ is not in lockdown}.
    \end{cases}
\end{equation}
Given a country that is ``not in lockdown'', we ascertain whether the country should transition to an ``in lockdown'' state based on their similarity to other countries that are ``in lockdown''; for country $i$ which is ``not in lockdown'', i.e.\ $\theta_t (i) = 0$, we calculate the distance  between it and other countries, $d_{ij} \forall j \in \{j | \theta(j) = 1 \}$.

There is also a global parameter, $p$, which denotes how many other countries an agent takes into consideration when evaluating its own average distance to the countries already in lockdown.
This parameter, $p$, has a value of $p \approx 18$ in model runs calibrated to the empirical data hinting realistic peer group size among countries.

Eventually agent $i$ evaluates the average of the $p$ countries with the least distance to themselves and if this average undercuts a country-specific threshold $s_i$, then the agent adopts a lockdown.
This process may be considered a `social' process since agents make the decision conditional on other agents' decisions. The condition for lockdown adoption based on these social factors is therefore:
\begin{equation}\label{third_equation}
\frac{1}{p} \sum_{j=1}^{p} d_{ij} < s_i \quad \textrm{where} \quad  s_i \in [0,1].
\end{equation}
The threshold $s_i$ is set through model calibration and such that it is proportional to the degree of democracy in a country. This aligns with the empirical results by \textcite{Sebhatu2020} who found that democratic countries are particularly sensitive to lockdown adoption through social mimicking. 

There is also an a-social adoption-mechanism. This distinction is similar to the classic product-diffusion model by \textcite{bass1969new} where the two mechanisms are called `innovation' (a-social) and `imitation' (social). This a-social mechanism basically represents agents taking initiative on their own independently of their peers and is important to represent early adopters. The a-social adoption process is also modelled based on a simple adoption-threshold $b_i$ which itself is set proportional to the square of the logarithm of population-density as well as the inverse of the degree of democracy (or in other words the anti-democracy). This proportionality captures the variables influencing the `base willingness to adopt' a lockdown in line with what \textcite{Sebhatu2020} ascertained statistically. Furthermore, we assume that this `base willingness to adopt' is slightly influenced by the overall number of countries in lockdown. However, we model this global influence via an exponential function such that it remains entirely insignificant at first and only makes a measurable impact when more than $90\%$ of all countries already have adopted, resulting in a slight push to the laggards who have not adopted due to pure similarity with other adopters. This represents a further pressure by global social majority but it is not related to any similarity measure or group affiliation, so therefore it does not belong to the social mechanism presented in equations \ref{first_equation} to \ref{third_equation}. We implement the a-social mechanism via a random draw from a uniform distribution at every time step. In precise terms, the a-social adoption condition then is: 
\begin{equation}\label{fourth_equation}
P(X < b_i)=b_i \quad \textrm{where} \quad  X {\sim} U[0,1] \quad \textrm{and} \quad b_i \in [0,1].
\end{equation}
The threshold  \(b_i\) is set the following way: 
\begin{equation}\label{fifth_equation}
b_i = P^2*\frac{1}{Y_i}* B \quad \textrm{with} \quad 0 \leq b_i \leq 1,
\end{equation}
where $B$ is a globally calibrated parameter, $Y_i$ the democracy index of agent $i$ normalized on the variable-average and $P$ is the logarithm of the population density again in line with the findings by \textcite{Sebhatu2020}.

\begin{figure}[ht]
\centering
 \includegraphics[width=11cm,  height=7cm]{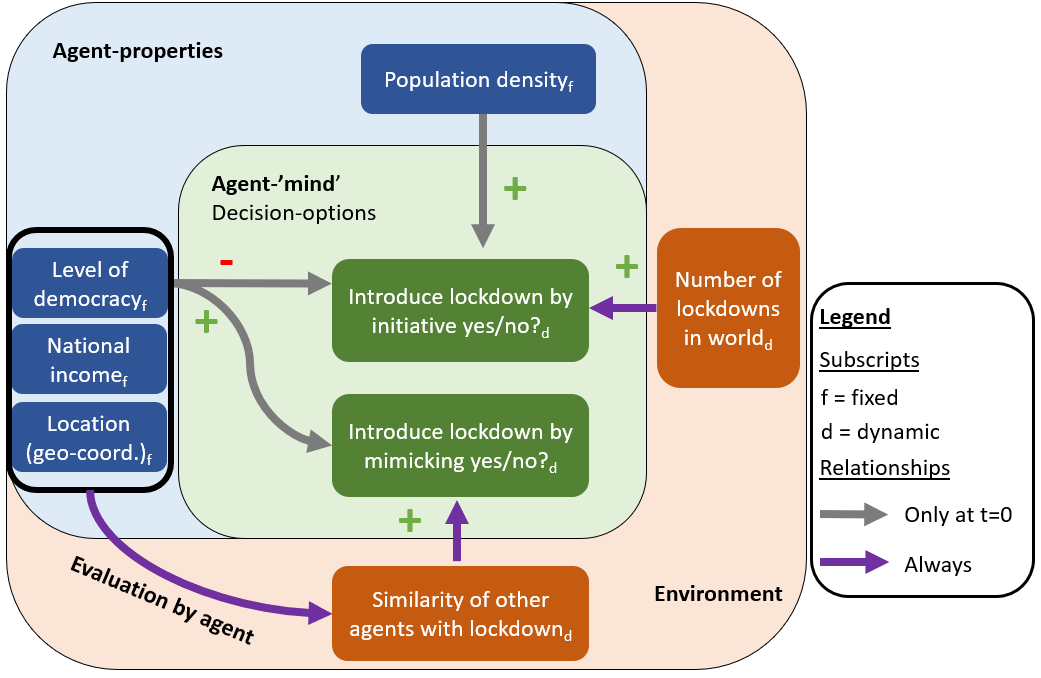}
  \caption{Model overview}
  \label{Figure 1: Model overview}
\end{figure}

Figure \ref{Figure 1: Model overview} summarizes the model in a high-level manner. There are three layers to the figure: The outer layer represents the environment in which agents are situated, the middle-layer represents the agent with all its properties, and the inner layer the `cognitive' layer of the agent's decision options. As described above, there are only two decision options (`adopt lockdown' or `do not adopt lockdown') and two decision mechanisms (`initiative' or through `peer pressure'). 
Finally, the model agent activation works in random yet sequential order.

\FloatBarrier

\subsection{Policy data description}\label{sec:policy_data}

To test and calibrate our model we employ data provided by Our World in Data \parencite{owidcoronavirus} that were originally gathered through the team of the Oxford COVID-19 Government Response Tracker \parencite{hale2021global}.
\textcite{hale2021global} not only gathered the original data but also demonstrated that almost all countries across the world ``ramped up'' their policies in the same 2-week window despite being affected by the virus to varying degrees. 
The data follows a variety of COVID-19 policy response measures on the national level, including school closures, workplace closures, face coverings, internal movement restrictions, international movement restrictions, public event cancellation and more. 

Each policy is measured on a categorical scale of usually three to five categories, with the exact scale depending on the indicator. The categorical levels represent the stringency of the policy. For example, in terms of school closures there are four levels (no measures, recommended, required only at some level, required at all levels). The indicators mostly ignore sub-national heterogeneity. If at least one sub-national region implemented the strictest measures, then the data classifies the entire country as having implemented this measure. This a substantial simplification and limitation. In Germany, for example, where school policies are generally the authority of the federal states, there was some heterogeneity. Although at a sub-national level the federal states possibly tended to mimic each other too, at the very least the policies were spatially correlated \parencite{fuchs2022covid}. Our goal is to model and describe (inter-)national policy diffusion, therefore we acknowledge this limitation but do not further deal with regional complexities.

\begin{figure}[ht]
\centering
 \includegraphics[width=12cm, height=5.5cm]{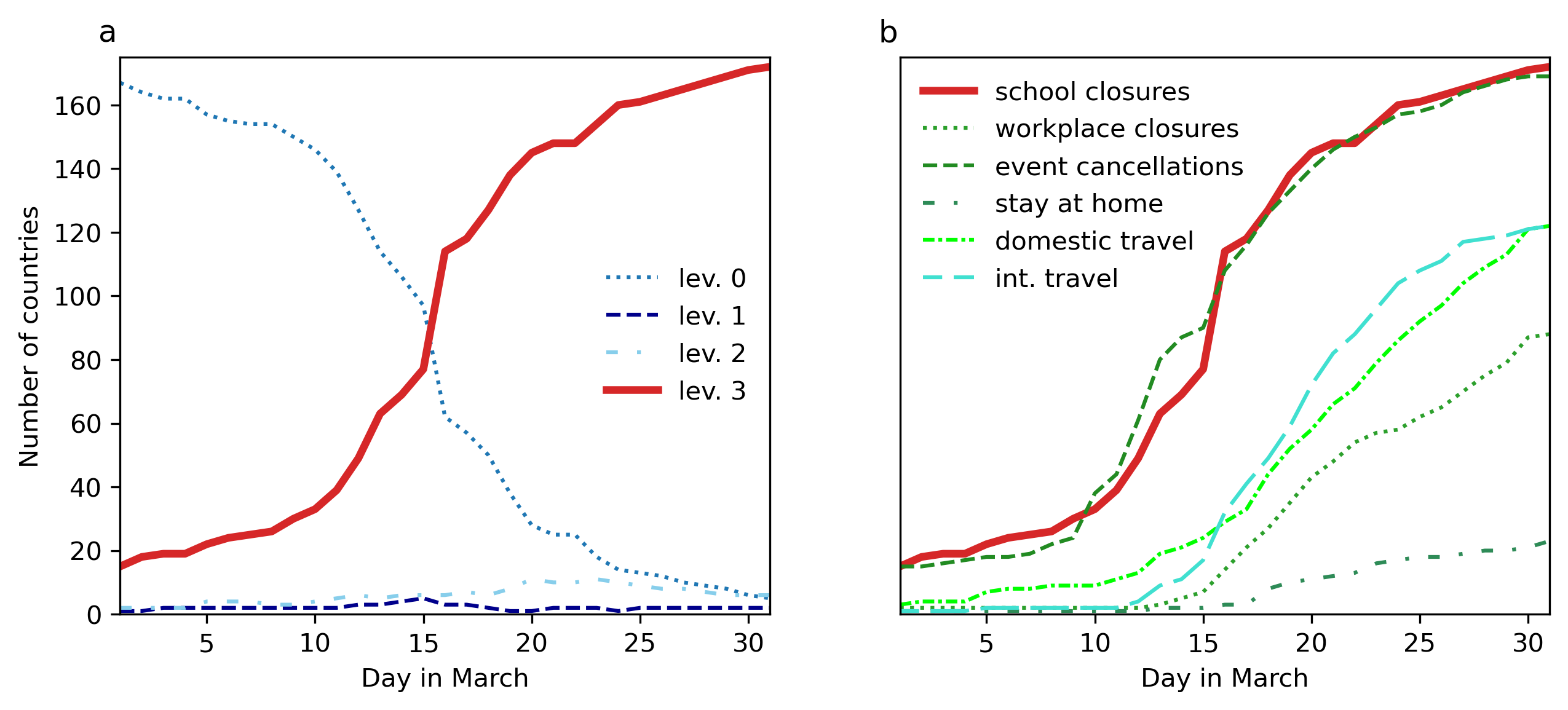}
  \caption{Data properties.}
  \label{Figure 2: Data properties}
  \medskip
\small
In this Figure we display data taken from \textcite{owidcoronavirus}.
\end{figure}

Based on the data, we can observe that for some policies countries did not implement intermediate stringency levels frequently but either went all-in or remained tolerant (of course this is partly influenced by how the measures' stringency has been interpreted in the first place). In Figure \ref{Figure 2: Data properties}(a), the diffusion of school closures (red solid line) across countries is depicted in terms of the four possible categories (level 0 to level 3). One can clearly observe the transition from level zero to level three without countries spending any time at level one or two. This implies that as a good first order approximation we might assume a binary policy choice (school closure or not). Secondly, Figure \ref{Figure 2: Data properties}(b) illustrates the diffusion of several policy measures taking on their highest level. Public event cancellations correlates very closely with school closures. Other measures do not follow school closures that closely but still align to a high degree. Partly this is because we only display the highest stringency level and for some metrics intermediate steps are more significant. For instance workplace closure correlates more closely with school closures and event cancellations if the second highest level ``required for some'' is also considered. In sum, all policies correlate substantially over time (Pearson's \(\rho > 0.9\) for all policies). Therefore, for modelling purposes, we assume that the system state ``lockdown'' is in line with the policy measure ``school closure'', which at the very least captures two policies (school closures and event cancellations) and exhibits substantial correlation with other measures.

\FloatBarrier

\subsection{Particle Filter description}

We implement a particle filter to constrain the evolution of our model to real-world observations as they emerge. A particle filter is essentially a genetic algorithm on the different simulation runs, which are called the particles, filtering out the ones that do not fit incoming data  well enough. Another way of defining a particle filter is that it is a step-wise algorithm to optimally estimate the posterior distribution of the state of a stochastic system, given a description of the system (the model) as well as real-time observations.

A particle filter assigns a weight to each model run based on a specified error metric which compares the model state to the observed system state. The filter can be thus denoted as the following set after \textcite{malleson2020}:
\begin{equation}\label{sixth_equation}
P_k = \Bigl\{ (x_k^i,w_k^i): i \in {1,...,N_p} \Bigr\}
\end{equation}
where \(N_p\) is the number of particles, \(x_k^i\) is the state vector of the i-th particle at the k-th observation, \(w_k^i\) is the
corresponding weight associated with particle \(i\) at observation \(k\), and the weights are subject to the condition \(\sum_{i=1}^{N_p} w_k^i = 1\).

Our model state can be specified by a vector with 164 cells (one for each country), each set to either zero (`not in lockdown') or to one (`in lockdown'). The observation state vector is the analog for the actual observation. Hence the difference between model estimation and observation for each particle \(i\), that is the error, can be computed as the fraction of countries to be estimated in their incorrect state \(c\), or simply:
\begin{equation}\label{seventh_equation}
e_k^i = 1 - \frac{c_k^i}{164}.
\end{equation}
The weights per particle filter at observation \(k\) are then proportional to the error squared \(w_k^i \propto (e_k^i)^2\). 

\begin{figure}[ht]
\centering
 \includegraphics[width=12cm,  height=4.0cm]{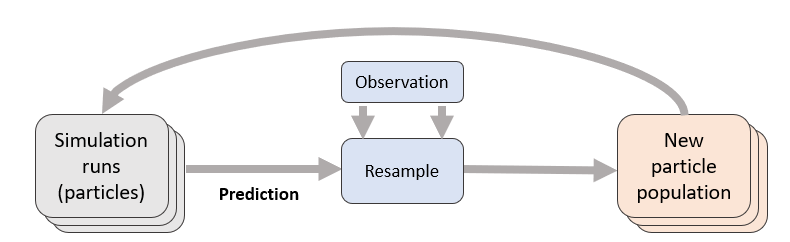}
  \caption{Particle Filter steps}
  \label{Figure 3: Particle Filter steps}
\end{figure}

After every reweighting procedure, a resampling of particles is undertaken to optimise the estimation of the system state. Here, Sequential Importance Resampling is used~\parencite{doucet_sequential_2000}. During this procedure, the weights are cumulatively counted, so they constitute a cumulative distribution function (CDF). This distribution of weights is compared against a uniformly random partition of the interval [0,1], constituting a uniformly random CDF. Then \(N_p\) points along the uniform distribution are selected, exactly at the mean step size of \(1/N_p\) and compared against the CDF of the weights at that particular point. For example, if there are 10 particles then the uniform CDF is evaluated at \(x = 0.1\), \(x = 0.2\) and so forth. Let us say at \(x = 0.1\) the uniform yields exactly \(y_u = 0.1\) and the weight distribution yields \(y_w = 0.2\). The uniform partition therefore makes a smaller step than the weight distribution. Then the weight of the particle is large, its error small and correspondingly the particle should be resampled. If on the other hand \(y_w = 0.05\) then the particle weight is less than the expected uniform average (which is 0.1) and thus the particle is discarded from the future particle population. Overall by conducting this procedure until 100\% of the uniform distribution are reached, it is very likely that particles with small weights are discarded because the `room' they make up in the cumulative weight distribution is very small. 
Figure~\ref{Figure 3: Particle Filter steps} depicts the particle procedure and its iterative nature. A certain number of particles are projected forward in time and then considering new observations, a new particle population is resampled and again this particle population is projected forward.

\FloatBarrier

\subsection{Experimental setups}

\subsubsection{Base run and calibration}\label{sec:base_run_calibration}

The first computational experiment we conduct is simply to fit the model to the data.
We calibrate the model parameters such that the diffusion curve of national lockdowns over time (see Figure \ref{Figure 2: Data properties}) is well approximated by the mean of an ensemble of 
simulations (where \(N = 100\)). We do minimize the distance of mean prediction and data but without specifying an exact loss function for this experiment because the added merit of the optimal calibration is small. We only require a model that captures the data relatively well before starting our data assimilation experiments. At this stage we measure the fit simply as correlation between mean prediction and data.  In this specific calibration, the probability of countries to adopt a lockdown on their own is set to 1\% and then further adjusted for each country by their degree of democracy and population density. It ranges from 0.0002 to 0.07 with most values around 0.01 and being roughly log-normally distributed (see supp-mat fig X). The social threshold is set to 0.13 and then additionally weighted by the degree of democracy in each country. It ranges from 0.02 to 0.25 and is approximately uniformly distributed in between (although there two notable modes near 0.075 and 0.15). The clique size that countries consider in their decision is set to 18 because it yields the most plausible shape of diffusion curves. 

Parameter-variations and more detailed explorations of the behaviour of the model can be found in the Supplementary Note 2. Table 1 provides a brief overview of the calibrated key parameters in the model.

\begin{table}[ht]
\centering
\caption{Parameters base run}
 \begin{tabular}{ | l | l |} 
  \hline
  Parameter name & Value \\ 
  \hline
  A-social threshold (global) & 0.01 \\ 

   Social threshold (global) & 0.13 \\

   Peer group size & 18 \\
     \hline
\end{tabular}
\end{table}

\FloatBarrier

\subsubsection{Particle filter calibration}

We test the particle filter in several ways. First we choose a medium case of the particle filter, that is a specific particle filter configuration with respect to the number of particles considered and the size of the data assimilation window and see how this compares to the model base run. For this purpose, we set the data assimilation window to five days and create 1000 particles. Subsequently we conduct two sensitivity analyses on the filter parameters. First, we test the influence of the number of particles on the filter performance, and second, we test the influence of the data assimilation window size, that is the frequency of applied filtering, on the filter performance. 

\section{Results}

\subsection{Experiment 1: Base run, calibration and validation}
\begin{figure}[ht]
\centering
  \includegraphics[width=14cm, height=6cm, center]{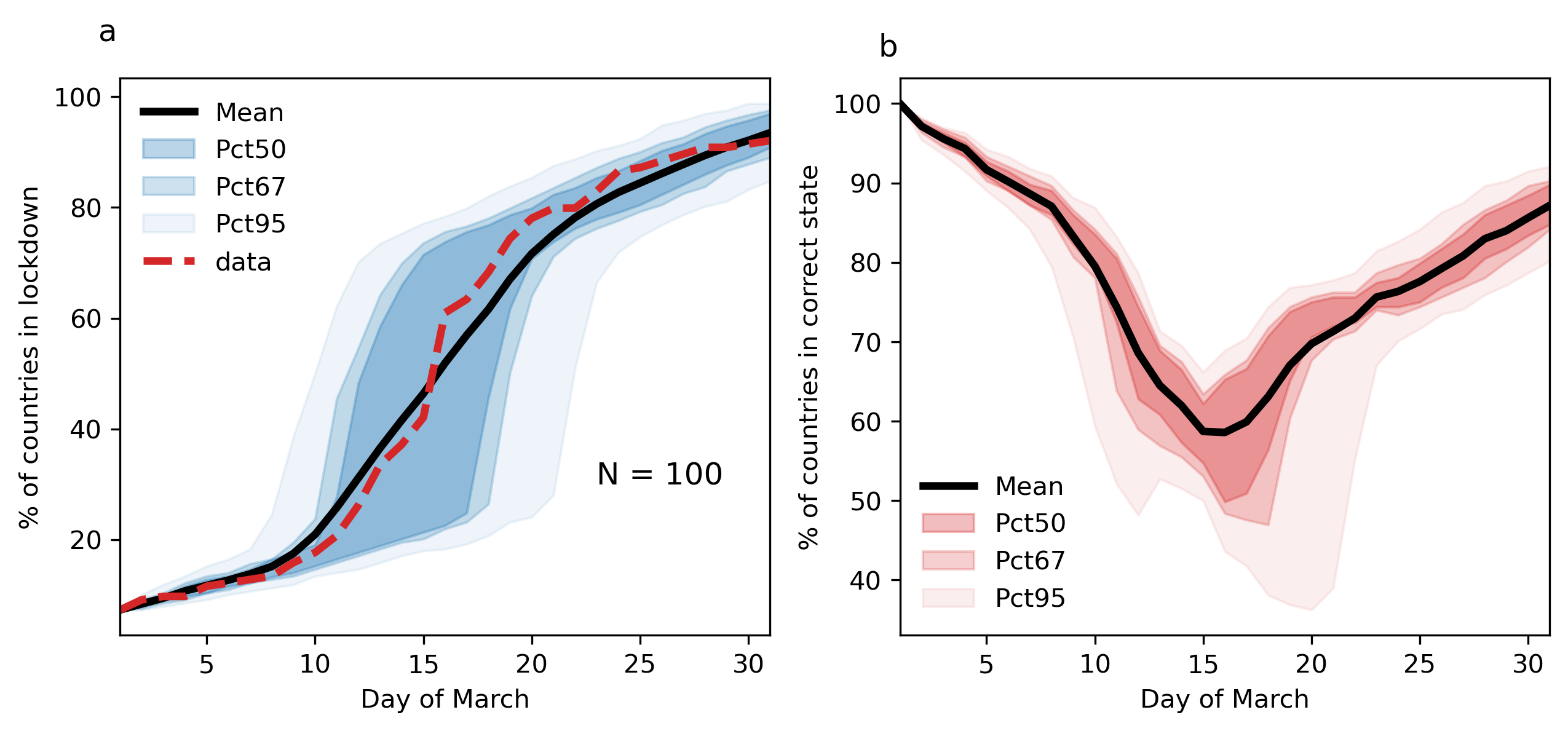}
   \caption{Base model ensemble run. (a) Percentage of countries that are in lockdown (`macro-results'). (b) Percentage of countries in the correct state (`micro-resuts').}
  \label{Figure 4: Base model ensemble run}
\end{figure}

We test the performance of the model without the support of a particle filter by employing a model ensemble of 100 runs with the above (Section~\ref{sec:base_run_calibration}) specified parameter configuration. Figure~\ref{Figure 4: Base model ensemble run} presents the `macro' results in panel (a) and the `micro' results in panel (b).

With respect to the `macro' results, the model only needs to estimate the total number of countries in lockdown. To this end it does very well, achieving a very high correlation between data and mean prediction of the model ($\rho > 0.99$). The model mean prediction (black solid line in Figure~\ref{Figure 4: Base model ensemble run}(a)) tracks the data (red dashed line) closely and the data are contained in the 95\% confidence interval. Importantly, the mean-squared error (MSE) between data and mean prediction is generally less than 10\% on the vertical scale of percentage of countries in lockdown. For instance, in a typical ensemble of 100 model runs (i.e. Figure \ref{Figure 4: Base model ensemble run}(a)) the maximum deviation between mean model prediction and data occurs around half way through the considered time period at the 16th of March. Here, the absolute deviation between the percentage of predicted countries in lockdown and actual countries in lockdown reaches on the order of 10\% (generally slightly less though).

\FloatBarrier

Variance and standard deviation are comparatively large however, even though only one parameter configuration is considered. The standard deviation of the ensemble reaches a maximum during the critical phase of the diffusion curve, roughly half way through March (at 23\% of all considered countries).

Considering the `micro' results (Figure \ref{Figure 4: Base model ensemble run}(b)), we now determine whether the model is able to predict the lockdown states of the \textit{individual counties}, rather than just predicting the total number of countries in lockdown. To do this we compare the the 164-dimensional vector (that has either a 0 or 1 in each cell representing whether each country is in lockdown or not) created by the model to that of the real observations. We find that the model performs worse, as expected, but still to a respectable degree. 
During the critical phase, mid March, the percentage of correctly estimated countries deteriorates to nearly 50\% only but recovers towards the end of the simulation. The variance of micro states estimates is generally larger towards the second half of March, and particularly high during the critical phase. This is because a number of simulations get the system state substantially wrong at around the 20th of March. Observing Figure \ref{Figure 4: Base model ensemble run}(a) again we see that the 95\% interval includes simulations that vastly underestimate the number of countries at that point in time. This is due to the stochasticity in the a-social adoption mechanism. If too few countries adopt a lockdown on their own, no mass-adoption is triggered because not enough countries find a sufficiently similar country that would have done the same. The importance of correctly capturing the behaviour in this `critical phase' highlights the need to update the model with current data, as is the subject of the next section.

\subsection{Experiment 2: Particle filter}

\subsubsection{Experiment 2.0: Particle filter compared to model base run}

\begin{figure}[ht]
\centering
  \includegraphics[width=14cm, height=6cm, center]{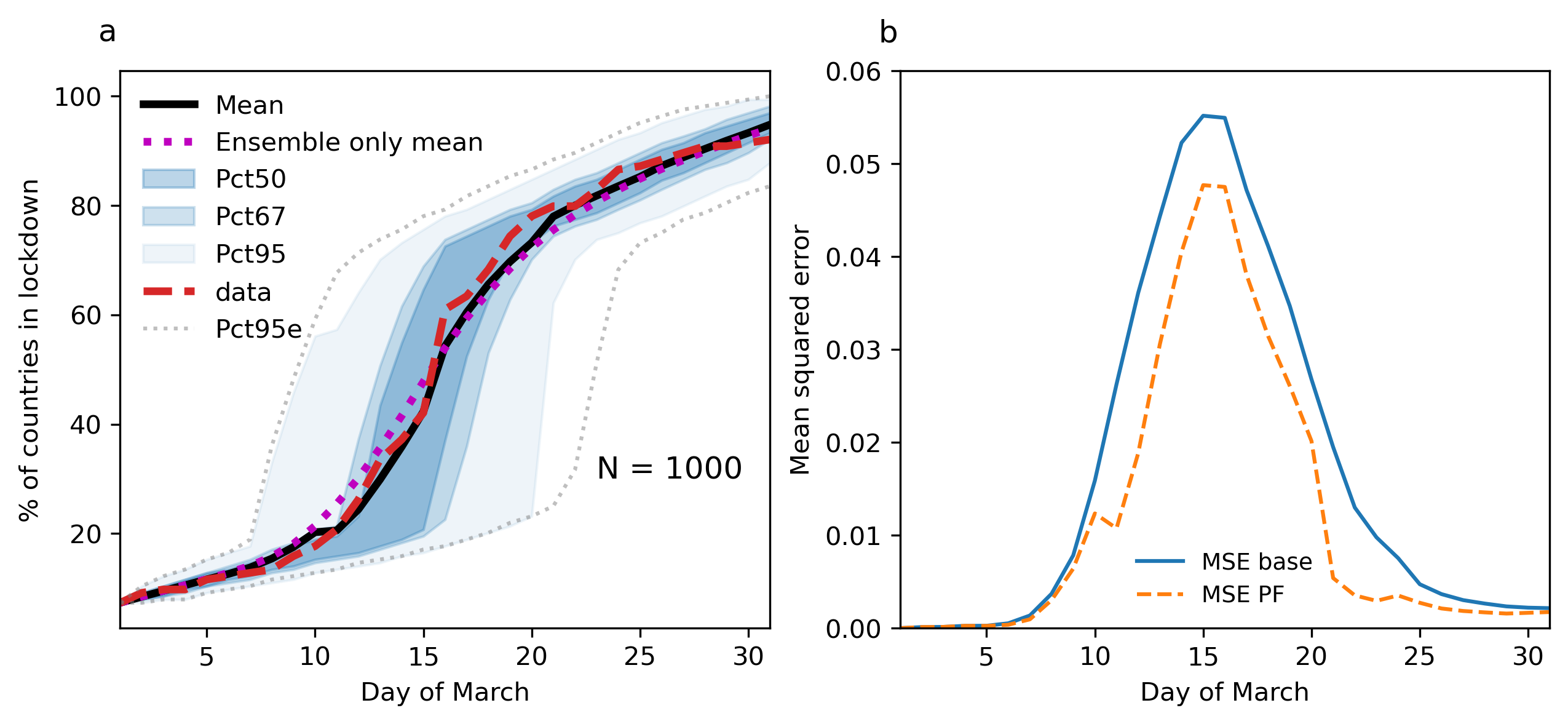}
   \caption{Particle filter compared to base model ensemble}
  \label{Figure 5: Particle filter compared to base model ensemble}
\end{figure}

In this experiment, we run both the base model and the model plus particle filter 1000 times. We re-sample particles at every fifth time step; that is, the models are confronted with observations on every fifth day of March 2020. The results are plotted in Figure \ref{Figure 5: Particle filter compared to base model ensemble}. The particle filter improves the fit of the mean model prediction to the data and lowers the variance in results. Hence it increases the reliability of prediction. In Figure \ref{Figure 5: Particle filter compared to base model ensemble}, the particle filtered model mean is plotted as solid dashed line and the previous `ensemble only' mean as magenta-colored dotted line. For comparison with the new 95\% confidence intervals, the previous ones are illustrated as dotted lines. In particular, the 50\% confidence interval is narrowed down and effectively halves along the temporal dimension (dark blue shaded region in Figures \ref{Figure 4: Base model ensemble run} and \ref{Figure 5: Particle filter compared to base model ensemble}). To quantify the particle performance better, we plot the mean-squared error (MSE) over time from the base model run compared to the MSE of the particle filter. At a maximum, the particle filter reduces the MSE by nearly 75\%. This happens roughly from 20th to 25th March. During the critical phase between 10th and 20th March, there is a substantial reduction of the MSE on the order of 10\% to 40\% depending on the exact time point considered.

\FloatBarrier

\subsubsection{Experiment 2.1: Number of particles sensitivity}\label{sec:experiment2.1}

In this experiment, we vary the number of particles $N$ along the exponential function $N(x) = 2^{x}$ with $x \in \mathbb{Z}$. The principal result is that the particle filter performs better than the model alone and this is independent of the particle numbers. The performance metric we have chosen to test is the mean-squared error (MSE) over time but summed up along the time axis. Since the time axis proceeds in unit steps of one day, this metric can be interpreted as numerical integral of the curves in Figure \ref{Figure 5: Particle filter compared to base model ensemble}(b). For lower particle numbers from  $2^{6} = 64$ to $2^{9} = 512$, we have conducted 20 test runs. This means, for instance, we ran 64 x 20 simulations in total. 
At those lower particle numbers, for example 64, the base model may still perform better in a number of cases. The distributions of iterations overlap. As the particle number increases, the distributions clearly separate and the particle filter reliably performs better than the model without particle filter. Hence, for computational efficiency, we have only conducted one iteration from \(2^{10} = 1024 \) particles onwards. Taking \(2^{12} = 4096 \) particles as a reference, the particle filter reduces the sum of the MSE over time by roughly 30\%, which is also the order of improvement in the other cases.

\begin{figure}[ht]
\centering
  \includegraphics[width=10cm, height=6.5cm, center] {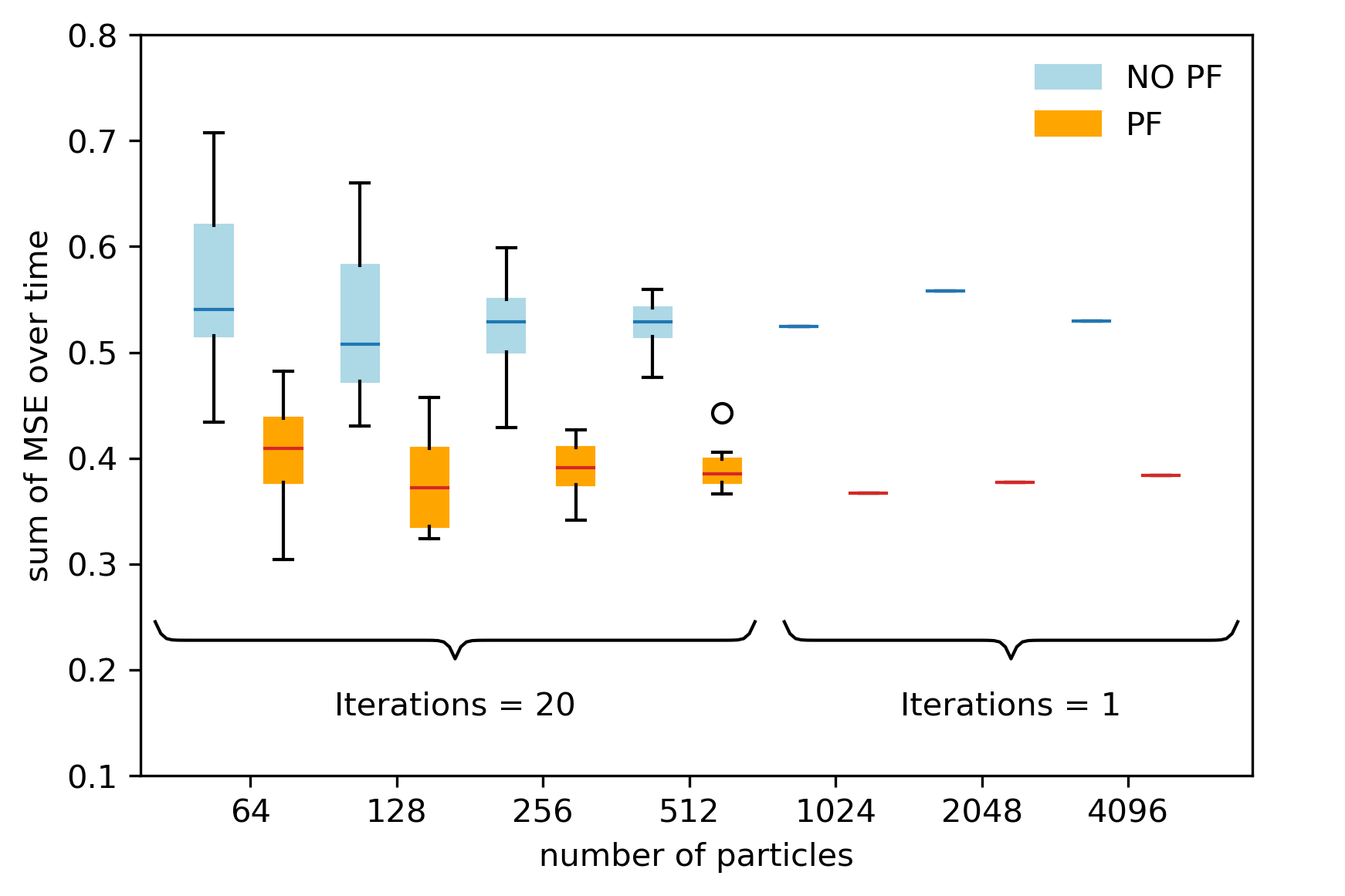}
   \caption{Experiment with the number of particles. In all cases the particle filter (`PF', orange) produces more accurate results than the random ensemble of models (`NO PF', blue).}
  \label{Figure 6: MSE_number_of_particles_exp.png}
\end{figure}

\FloatBarrier

\subsubsection{Experiment 2.2: Data assimilation window sensitivity}

Since we established in Section~\ref{sec:experiment2.1} that at a minimum on the order of $2^{9}$ particles are required to get a reliable filter, here we fix the particle quantity at 1000 but vary the filtering frequency. We test six such frequencies: (i) no filtering; (ii) filtering at every 15th time step, so twice; (iii) at every 10th time step, so three times; (iv) at every 5th time step, so 6 times; (v) at every 2nd time step so 15 times; and lastly (vi) at every time step, so 31 times. 

\begin{figure}[ht]
\centering
  \includegraphics[width=14cm, height=6cm, center]{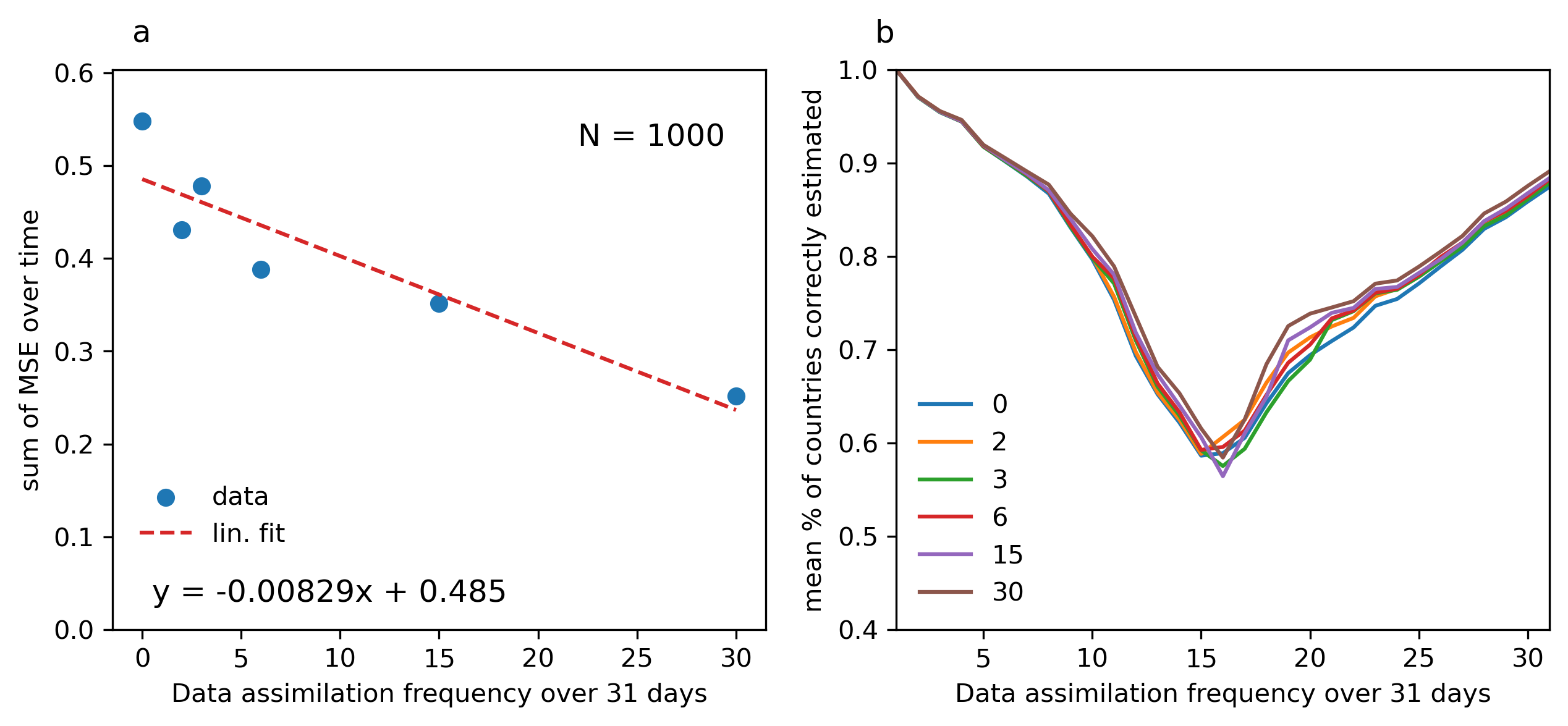}
   \caption{Filtering frequency effect on model performance}
  \label{Figure 7: Filtering frequency effect on model performance}
\end{figure}

The results indicate a linear relationship between the filtering frequency and the MSE of the aggregate diffusion curve. For every additional five days of filtering, the sum of the MSE over time is reduced by roughly 10\% as compared to no filtering. Thus, the aggregate estimation of how many countries out of the global total are in lockdown is significantly improved through the particle filter. The impact on the micro-accuracy is more marginal however. There is no substantial improvement in exactly knowing what country switches when. This means that through the particle filter we do not gain better information on when exactly specific countries switch to lockdown. Increased filtering does still lead to a slightly better micro-performance but not substantially, as seen in Figure~\ref{Figure 7: Filtering frequency effect on model performance}(b). This is likely due to a model limitation rather than a filter limitation. The model does not produce trajectories which perform substantially better than being 60\% correct during the critical transition phase from roughly 10th to 20th March (see Figure \ref{Figure 4: Base model ensemble run}(b)). The model ensemble simply does not include trajectories that perform substantially better than the average. Consequently, the particle filter cannot find a much improved optimum compared to the average prediction quality.

\FloatBarrier

\section{Discussion and conclusions}

\subsection{Interpretation and generalizability}

We have presented a model of the COVID-19 policy response diffusion across nations and optimized the model estimation in `real time' by employing a particle filter.
The model is successful in reproducing the observed diffusion pattern based on peer mimicry as an interactive mechanism alone, if calibrated to the data. 
The model comes with limitations however. We have not tested diffusion proceeding over diverse periods of time, for instance over a year instead of a month. Therefore, the applicability of the model so far is limited to the presented case-study and the robustness of the model needs to be assessed with care. While capturing the particular diffusion phenomenon presented here, it is unclear whether the mimicry mechanism can be generalized at all to other case studies and in particular to slower international policy diffusion patterns, such as the spread of environmental reforms. The COVID-19 policy tipping point remains an extraordinary situation, enabled by great uncertainty and a sense of danger. A model appealing to the more general case likely must draw from a greater diversity of diffusion mechanisms, such as those discussed in the literature review Section~\ref{sec:policy_diffusion}. Moreover, even in the particular case of peer mimicry alone, the dimensions considered (income, degree of democracy and geographical position) are far from the only plausible ones. We could have considered, for instance, trade relationships or international connectivity of airports or proceeded in more general terms, employing variables such as ``cultural exchange'' or ``cultural proximity'' (whose exact operationalization is challenging). We defer the exploration of further mimicry drivers to future extensions. Ultimately, we have not attempted to construct the best possible national policy diffusion model but to come up with a working first order heuristic that can be optimized by data assimilation. In a real-world and real-time application, let us say another pandemic or another environmental and geopolitical crisis, an ABM could make predictions before an actual great shift in global has happened, and be ``constrained by the data'' as time goes on. 

In terms of predicting rapid transitions, we by no means claim to have found a better approach than other prediction approaches, such as for example, Early Warning Signals (EWS) of tipping points. Early Warning Signals is an approach inferring generic signs of tipping points in time series data, such as unusual fluctuations and auto-correlations \parencite{scheffer2009early}. Time series data however might not be always available. In our case study, there would have been no sufficiently long time series on countries switching in- and out of lockdowns. And even with the help of EWS it remains difficult to detect transitions. Our agent-based modelling approach, in combination with data assimilation, presents a complementary perspective that combines theory and data for prediction instead of only scanning for statistical anomalies.

Perhaps the most generalizable conclusion from our research is that we delivered a proof-of-concept that social and even political agent-based diffusion models can be combined with data assimilation. Social diffusion contains a large class of phenomena after all, ranging from product-diffusion to the spread of opinion and information in social media. Especially the latter often unfold rapidly and in real-time, on the order of seconds and minutes, with data being made available constantly. And there are numerous ABMs trying to capture these processes \parencite{kvasnivcka2014viral, chen2019agent}. Hence, the combination of diffusion oriented ABMs and data assimilation likely offers promising research opportunities.

\subsection{Outlook}

As discussed above, the model performs best at the aggregate level but less so at modelling the micro-level, because the model does not produce particles that are much better than the mean quality of prediction and the particle filter hence can not find any optimum that is substantially beyond. In future work then, it must be imperative to improve the model rather than only the data assimilation algorithms. 

Besides, a more nuanced model likely opens manifold new opportunities for data assimilation. We have so far reduced the model state to a binary variable (lockdown or no lockdown), even though the reality of course was far more complex than that. Countries actually implemented diverse sets of policies with varying degrees over specific issues (school closures for instance). Therefore, a first opportunity would be to consider a higher dimensional system state -- with more categorical variables operationalizing the actual lockdown policies, possibly even numerical variables. A more complex system state however might go beyond what the particle filter can reasonably optimize, since with increasing system complexity, a particle filter becomes rapidly more computational expensive, as evidenced by attempts to apply the method to more complex agent-based models~\parencite{malleson2020, ternes2022data}. Therefore other data assimilation algorithms that do not rely on a plethora of particles, like the Extended Kalman Filter, could be tested next, together with extensions of the model.

\section*{Acknowledgements}

This project has received funding from the European Research Council (ERC) under the European Union’s Horizon 2020 research and innovation programme (grant agreement No. 757455).

\section*{Code and data accessibility}

All code and data relating to this study is open-access at \url{https://github.com/eeyouol/covpol}.

\section*{Author Biographies}

Dr Oswald is a Postdoctoral Researcher in agent-based modelling and data-assimilation at the School of Geography, University of Leeds. He has a PhD in Ecological Economics and a M.Sc. in International Sustainable Development. His research covers inequality, climate change, complex systems and computational social science. Recently, he has taken interest in the merger of data science and complexity science and its applications to global-scale challenges such as climate change, pandemics and inequality.

Dr Suchak is a Research Fellow in Transport-Health Interactions at the Institute for Transport Studies, University of Leeds. He has a PhD in Geography, Masters degrees in Aerospace Engineering, Mathematics and Data Analytics, and an undergraduate degree in Physics. His research focuses on the development of novel methods, particularly in relation to Agent-Based Models and the simulation of social systems. His recent work has focused on developing methods for real-time pedestrian simulation, and using a variety of simulation methods to explore the health impacts of changes in transport choices.

Dr Malleson is a Professor of Spatial Science at the Centre for Spatial Analysis and Policy at the School of Geography, University of Leeds, UK. He has a PhD in Geography and undergraduate degrees in Computer Science (BSc) and Multidiciplinary Informatics (MSc). Most of his research focuses on the development of spatial computer models that help to understand and explain social phenomena. He has a particular interest in simulations of crime patterns, and in models that can be used to describe the flows of people around cities. More recently, he has become in interested in how `big data', agent-based modelling, and smart cities initiatives can be used to better understand the daily dynamics of cities and reduce the impacts of phenomena such as pollution or crime.

\printbibliography

@article{wang_data_2015,
  title = {Data Assimilation in Agent Based Simulation of Smart Environments Using Particle Filters},
  author = {Wang, Minghao and Hu, Xiaolin},
  year = {2015},
  journal = {Simulation Modelling Practice and Theory},
  volume = {56},
  pages = {36--54},
  issn = {1569-190X},
  doi = {10.1016/j.simpat.2015.05.001},
}

@inproceedings{lueck_who_2019,
  title = {Who {{Goes There}}? {{Using}} an {{Agent-based Simulation}} for {{Tracking Population Movement}}},
  booktitle = {Winter {{Simulation Conference}}, {{Dec}} 8 - 11, 2019.},
  author = {Lueck, Jordan and Rife, Jason H. and Swarup, Samarth and Uddin, Nassin},
  year = {2019},
  address = {{National Harbor, MD, USA}},
}

@article{doucet_sequential_2000,
  title = {On Sequential {{Monte Carlo}} Sampling Methods for {{Bayesian}} Filtering},
  author = {Doucet, Arnaud and Godsill, Simon and Andrieu, Christophe},
  year = {2000},
  journal = {Statistics and Computing},
  volume = {10},
  number = {3},
  pages = {197--208},
  issn = {09603174},
  doi = {10.1023/A:1008935410038}
}

@article{Sebhatu2020,
   author = {Abiel Sebhatu and Karl Wennberg and Stefan Arora-Jonsson and Staffan I Lindberg},
   doi = {10.1073/pnas.2010625117/-/DCSupplemental},
   issue = {35},
   journal = {Proceedings of the National Academy of Sciences of the United States of America},
   title = {Explaining the homogeneous diffusion of COVID-19 nonpharmaceutical interventions across heterogeneous countries},
   volume = {117},
   url = {www.pnas.org/cgi/doi/10.1073/pnas.2010625117},
   year = {2020},
}

@book{Bercovitch2009,
 ISBN = {9780472070626},
 URL = {http://www.jstor.org/stable/10.3998/mpub.106467},
 author = {Jacob Bercovitch and Richard Jackson},
 publisher = {University of Michigan Press},
 title = {Conflict Resolution in the Twenty-first Century: Principles, Methods, and Approaches},
 urldate = {2022-08-29},
 year = {2009}
}

@article{Morse2012,
   author = {Stephen S Morse and Jonna A K Mazet and Mark Woolhouse and Colin R Parrish and Dennis Carroll and William B Karesh and Carlos Zambrana-Torrelio and Ian Lipkin and Peter Daszak},
   journal = {The Lancet},
   title = {Prediction and prevention of the next pandemic zoonosis},
   volume = {380},
   url = {www.thelancet.com},
   year = {2012},
}

@article{owidcoronavirus,
    author = {Hannah Ritchie and Edouard Mathieu and Lucas Rodés-Guirao and Cameron Appel and Charlie Giattino and Esteban Ortiz-Ospina and Joe Hasell and Bobbie Macdonald and Diana Beltekian and Max Roser},
    title = {Coronavirus Pandemic (COVID-19)},
    journal = {Our World in Data},
    year = {2020},
    note = {https://ourworldindata.org/coronavirus}
}

@article{Shukla2022,
   author = {Priyadarshi R Shukla and Jim Skea and Andy Reisinger and Raphael Slade and Roger Fradera and Minal Pathak and Alaa Al and Khourdajie Malek and Belkacemi Renée Van Diemen and Apoorva Hasija and Géninha Lisboa and Sigourney Luz and Juliette Malley and David Mccollum and Shreya Some},
   isbn = {9789291691609},
   title = {Climate Change 2022 Mitigation of Climate Change Working Group III Contribution to the Sixth Assessment Report of the Intergovernmental Panel on Climate Change Summary for Policymakers Edited by},
   url = {www.ipcc.ch},
   year = {2022},
}

@article{Daly2020,
   author = {Mary Daly and Bernhard, Ebbinghaus and Lukas Lehner and Marek Naczyk and Tim Vlandas},
   title = {Oxford Supertracker: The Global Directory for COVID Policy Trackers and Surveys, Department of Social Policy and Intervention},
   url = {https://supertracker.spi.ox.ac.uk/about/},
   year = {2020},
}

@article{Chetty2020,
   author = {Raj Chetty and John N. Friedman and Nathaniel Hendren and Michael Stepner and The Opportunity Insights Team},
   institution = {NBER WORKING PAPER SERIES},
   title = {How did Covid-19 and stabilization policies affect spending and employment? A new real-time economic tracker based on private sector data},
   url = {http://www.nber.org/papers/w27431},
   year = {2020},
}

@article{kiesling2012agent,
  title={Agent-based simulation of innovation diffusion: a review},
  author={Kiesling, Elmar and G{\"u}nther, Markus and Stummer, Christian and Wakolbinger, Lea M},
  journal={Central European Journal of Operations Research},
  volume={20},
  number={2},
  pages={183--230},
  year={2012},
  publisher={Springer}
}

@article{bohlmann2010effects,
  title={The effects of market network heterogeneity on innovation diffusion: An agent-based modeling approach},
  author={Bohlmann, Jonathan D and Calantone, Roger J and Zhao, Meng},
  journal={Journal of Product Innovation Management},
  volume={27},
  number={5},
  pages={741--760},
  year={2010},
  publisher={Wiley Online Library}
}

@article{xiao2016forecasting,
  title={Forecasting new product diffusion with agent-based models},
  author={Xiao, Yu and Han, Jingti},
  journal={Technological Forecasting and Social Change},
  volume={105},
  pages={167--178},
  year={2016},
  publisher={Elsevier}
}

@article{lorig2021agent,
  title={Agent-based social simulation of the COVID-19 pandemic: A systematic review},
  author={Lorig, Fabian and Johansson, Emil and Davidsson, Paul},
  journal={JASSS: Journal of Artificial Societies and Social Simulation},
  volume={24},
  number={3},
  year={2021},
  publisher={JASSS}
}

@article{kerr2021covasim,
  title={Covasim: an agent-based model of COVID-19 dynamics and interventions},
  author={Kerr, Cliff C and Stuart, Robyn M and Mistry, Dina and Abeysuriya, Romesh G and Rosenfeld, Katherine and Hart, Gregory R and N{\'u}{\~n}ez, Rafael C and Cohen, Jamie A and Selvaraj, Prashanth and Hagedorn, Brittany and others},
  journal={PLOS Computational Biology},
  volume={17},
  number={7},
  pages={e1009149},
  year={2021},
  publisher={Public Library of Science San Francisco, CA USA}
}

@article{cuevas2020agent,
  title={An agent-based model to evaluate the COVID-19 transmission risks in facilities},
  author={Cuevas, Erik},
  journal={Computers in biology and medicine},
  volume={121},
  pages={103827},
  year={2020},
  publisher={Elsevier}
}

@article{kai2020universal,
  title={Universal masking is urgent in the COVID-19 pandemic: SEIR and agent based models, empirical validation, policy recommendations},
  author={Kai, De and Goldstein, Guy-Philippe and Morgunov, Alexey and Nangalia, Vishal and Rotkirch, Anna},
  journal={arXiv preprint arXiv:2004.13553},
  year={2020}
}

@article{ternes2022data,
  title={Data assimilation and agent-based modelling: towards the incorporation of categorical agent parameters},
  author={Ternes, Patricia and Ward, Jonathan A and Heppenstall, Alison and Kumar, Vijay and Kieu, Le-Minh and Malleson, Nick},
  journal={Open Research Europe},
  volume={1},
  year={2022},
  publisher={F1000Research}
}

@article{malleson2020,
   title = {Simulating Crowds in Real Time with Agent-Based Modelling and a Particle Filter},
   author = {Malleson, Nick and Minors, Kevin and Kieu, Le-Minh and Ward, Jonathan A. and West, Andrew and Heppenstall, Alison},
   journal = {Journal of Artificial Societies and Social Simulation},
   ISSN = {1460-7425},
   volume = {23},
   number = {3},
   pages = {3},
   year = {2020},
   URL = {http://jasss.soc.surrey.ac.uk/23/3/3.html},
   DOI = {10.18564/jasss.4266},
}

@article{simmons_dobbin_garrett_2006, title={Introduction: The International Diffusion of Liberalism}, volume={60}, DOI={10.1017/S0020818306060267}, number={4}, journal={International Organization}, publisher={Cambridge University Press}, author={Simmons, Beth A. and Dobbin, Frank and Garrett, Geoffrey}, year={2006}, pages={781–810}}

@article{zhang2019multiple,
  title={Multiple mechanisms of policy diffusion in China},
  author={Zhang, Youlang and Zhu, Xufeng},
  journal={Public Management Review},
  volume={21},
  number={4},
  pages={495--514},
  year={2019},
  publisher={Taylor \& Francis}
}

@article{baybeck2011strategic,
  title={A strategic theory of policy diffusion via intergovernmental competition},
  author={Baybeck, Brady and Berry, William D and Siegel, David A},
  journal={The Journal of Politics},
  volume={73},
  number={1},
  pages={232--247},
  year={2011},
  publisher={Cambridge University Press New York, USA}
}

@article{gilardi2016four,
  title={Four ways we can improve policy diffusion research},
  author={Gilardi, Fabrizio},
  journal={State Politics \& Policy Quarterly},
  volume={16},
  number={1},
  pages={8--21},
  year={2016},
  publisher={SAGE Publications Sage CA: Los Angeles, CA}
  }

@article{shipan2008mechanisms,
  title={The mechanisms of policy diffusion},
  author={Shipan, Charles R and Volden, Craig},
  journal={American journal of political science},
  volume={52},
  number={4},
  pages={840--857},
  year={2008},
  publisher={Wiley Online Library}
}

@article{zimm2021improving,
  title={Improving the understanding of electric vehicle technology and policy diffusion across countries},
  author={Zimm, Caroline},
  journal={Transport Policy},
  volume={105},
  pages={54--66},
  year={2021},
  publisher={Elsevier}
}

@article{dobbin2007global,
  title={The global diffusion of public policies: Social construction, coercion, competition, or learning?},
  author={Dobbin, Frank and Simmons, Beth and Garrett, Geoffrey},
  journal={Annu. Rev. Sociol.},
  volume={33},
  pages={449--472},
  year={2007},
  publisher={Annual Reviews}
}

@article{brooks2004role,
  title={What was the role of international financial institutions in the diffusion of social security reform in Latin America?},
  author={Brooks, Sarah M},
  journal={Learning from foreign models in Latin American policy reform},
  pages={53--80},
  year={2004},
  publisher={Woodrow Wilson Center Press and Johns Hopkins University Press Washington, DC}
}

@article{gautier2021transnational,
  title={Transnational networks’ contribution to health policy diffusion: a mixed method study of the performance-based financing community of practice in Africa},
  author={Gautier, Lara and De Allegri, Manuela and Ridde, Val{\'e}ry},
  journal={International journal of health policy and management},
  volume={10},
  number={6},
  pages={310},
  year={2021},
  publisher={Kerman University of Medical Sciences}
}

@article{shen2014cross,
  title={Cross-national diffusion of mental health policy},
  author={Shen, Gordon C},
  journal={International journal of health policy and management},
  volume={3},
  number={5},
  pages={269},
  year={2014},
  publisher={Kerman University of Medical Sciences}
}

@article{lundgren2020emergency,
  title={Emergency powers in response to COVID-19: Policy Diffusion, democracy, and preparedness},
  author={Lundgren, Magnus and Klamberg, Mark and Sundstr{\"o}m, Karin and Dahlqvist, Julia},
  journal={Nordic Journal of Human Rights},
  volume={38},
  number={4},
  pages={305--318},
  year={2020},
  publisher={Taylor \& Francis}
}

@article{mistur2020contagious,
  title={Contagious COVID-19 policies: Policy diffusion during times of crisis},
  author={Mistur, Evan M and Givens, John Wagner and Matisoff, Daniel C},
  journal={Review of Policy Research},
  year={2020},
  publisher={Wiley Online Library}
}

@article{li2017agent,
  title={Agent-based modelling approach for multidimensional opinion polarization in collective behaviour},
  author={Li, Jin and Xiao, Renbin},
  journal={Journal of Artificial Societies and Social Simulation},
  volume={20},
  number={2},
  year={2017},
  publisher={JASSS}
}

@article{leifeld2014polarization,
  title={Polarization of coalitions in an agent-based model of political discourse},
  author={Leifeld, Philip},
  journal={Computational Social Networks},
  volume={1},
  number={1},
  pages={1--22},
  year={2014},
  publisher={SpringerOpen}
}

@article{rapaport2009puzzle,
  title={The Puzzle of the Diffusion of Central-Bank Independence Reforms: Insights from an Agent-Based Simulation},
  author={Rapaport, Orit and Levi-Faur, David and Miodownik, Dan},
  journal={Policy Studies Journal},
  volume={37},
  number={4},
  pages={695--716},
  year={2009},
  publisher={Wiley Online Library}
}

@article{kingdon1984agendas,
  title={Agendas, alternatives, and public policies},
  author={Kingdon, John W and Stano, Eric},
  volume={45},
  year={1984},
  publisher={Little, Brown Boston}
}

@article{braun2006taking,
  title={Taking ‘Galton's problem’seriously: Towards a theory of policy diffusion},
  author={Braun, Dietmar and Gilardi, Fabrizio},
  journal={Journal of theoretical politics},
  volume={18},
  number={3},
  pages={298--322},
  year={2006},
  publisher={Sage Publications London, Thousand Oaks, CA and New Delhi}
}

@article{luyet2011policy,
  title={Policy diffusion: An agent-based approach},
  author={Luyet, St{\'e}phane},
  year={2011},
  school={Universit{\'e} de Lausanne, Facult{\'e} des sciences sociales et politiques}
}

@article{ring2014agent,
  title={An agent-based model of international norm diffusion},
  author={Ring, Jonathan},
  journal={University of Iowa, Iowa City},
  volume={222},
  year={2014}
}

@article{o1998diffusion,
  title={The diffusion of democracy, 1946--1994},
  author={O'loughlin, John and Ward, Michael D and Lofdahl, Corey L and Cohen, Jordin S and Brown, David S and Reilly, David and Gleditsch, Kristian S and Shin, Michael},
  journal={Annals of the Association of American Geographers},
  volume={88},
  number={4},
  pages={545--574},
  year={1998},
  publisher={Taylor \& Francis}
}

@article{simmons2004globalization,
  title={The globalization of liberalization: Policy diffusion in the international political economy},
  author={Simmons, Beth A and Elkins, Zachary},
  journal={American political science review},
  volume={98},
  number={1},
  pages={171--189},
  year={2004},
  publisher={Cambridge University Press}
}

@article{owiddemocracy,
    author = {Bastian Herre and Max Roser},
    title = {Democracy},
    journal = {Our World in Data},
    year = {2013},
    note = {https://ourworldindata.org/democracy}
}

@article{tews2003diffusion,
  title={The diffusion of new environmental policy instruments 1},
  author={Tews, Kerstin and Busch, Per-Olof and J{\"o}rgens, Helge},
  journal={European journal of political research},
  volume={42},
  number={4},
  pages={569--600},
  year={2003},
  publisher={Wiley Online Library}
}

@article{barabasi1999emergence,
  title={Emergence of scaling in random networks},
  author={Barab{\'a}si, Albert-L{\'a}szl{\'o} and Albert, R{\'e}ka},
  journal={science},
  volume={286},
  number={5439},
  pages={509--512},
  year={1999},
  publisher={American Association for the Advancement of Science}
}

@article{batista2019migrant,
  title={Do migrant social networks shape political attitudes and behavior at home?},
  author={Batista, Catia and Seither, Julia and Vicente, Pedro C},
  journal={World Development},
  volume={117},
  pages={328--343},
  year={2019},
  publisher={Elsevier}
}

@article{kandiah2012pagerank,
  title={PageRank model of opinion formation on social networks},
  author={Kandiah, Vivek and Shepelyansky, Dima L},
  journal={Physica A: Statistical Mechanics and its Applications},
  volume={391},
  number={22},
  pages={5779--5793},
  year={2012},
  publisher={Elsevier}
}

@article{lu2021big,
  title={Big data-drive agent-based modeling of online polarized opinions},
  author={Lu, Peng and Zhang, Zhuo and Li, Mengdi},
  journal={Complex \& Intelligent Systems},
  volume={7},
  number={6},
  pages={3259--3276},
  year={2021},
  publisher={Springer}
}

@article{kreindler2014rapid,
  title={Rapid innovation diffusion in social networks},
  author={Kreindler, Gabriel E and Young, H Peyton},
  journal={Proceedings of the National Academy of Sciences},
  volume={111},
  number={supplement\_3},
  pages={10881--10888},
  year={2014},
  publisher={National Acad Sciences}
}

@article{tang2022data,
  title={Data assimilation with agent-based models using Markov chain sampling},
  author={Tang, Daniel and Malleson, Nick},
  journal={arXiv preprint arXiv:2205.01616},
  year={2022}
}

@article{ward2016dynamic,
  title={Dynamic calibration of agent-based models using data assimilation},
  author={Ward, Jonathan A and Evans, Andrew J and Malleson, Nicolas S},
  journal={Royal Society open science},
  volume={3},
  number={4},
  pages={150703},
  year={2016},
  publisher={The Royal Society}
}

@article{axtell2022agent,
  title={Agent-based modeling in economics and finance: Past, present, and future},
  author={Axtell, Robert L and Farmer, J Doyne},
  journal={Journal of Economic Literature},
  year={2022},
  publisher={American Economic Association}
}

@article{kalnay2003atmospheric,
  title={Atmospheric modeling, data assimilation and predictability},
  author={Kalnay, Eugenia},
  year={2003},
  publisher={Cambridge university press}
}

@book{gatti2018agent,
  title={Agent-based models in economics: A toolkit},
  author={Gatti, Domenico Delli and Fagiolo, Giorgio and Gallegati, Mauro and Richiardi, Matteo and Russo, Alberto},
  year={2018},
  publisher={Cambridge University Press}
}

@article{CLAY2021102386,
title = {Real-time agent-based crowd simulation with the Reversible Jump Unscented Kalman Filter},
journal = {Simulation Modelling Practice and Theory},
volume = {113},
pages = {102386},
year = {2021},
issn = {1569-190X},
doi = {https://doi.org/10.1016/j.simpat.2021.102386},
url = {https://www.sciencedirect.com/science/article/pii/S1569190X21000939},
author = {Robert Clay and Jonathan A. Ward and Patricia Ternes and Le-Minh Kieu and Nick Malleson}
}

@article{hu2022data,
  title={Data Assimilation For Simulation-Based Real-Time Prediction/Analysis},
  author={Hu, Xiaolin},
  booktitle={2022 Annual Modeling and Simulation Conference (ANNSIM)},
  pages={404--415},
  year={2022},
  organization={IEEE}
}

@article{hale2021global,
  title={A global panel database of pandemic policies (Oxford COVID-19 Government Response Tracker)},
  author={Hale, Thomas and Angrist, Noam and Goldszmidt, Rafael and Kira, Beatriz and Petherick, Anna and Phillips, Toby and Webster, Samuel and Cameron-Blake, Emily and Hallas, Laura and Majumdar, Saptarshi and others},
  journal={Nature human behaviour},
  volume={5},
  number={4},
  pages={529--538},
  year={2021},
  publisher={Nature Publishing Group}
}

@article{worldbankGDP,
         title  =   {GDP, PPP (constant 2017 international \$)},
         author =   {{World Bank}},
         year   =   {2022},
         url = {https://data.worldbank.org/indicator/NY.GDP.MKTP.PP.KD}
}

@article{economistDemocracy2019,
  title={Democracy Index 2019 A year of democratic setbacks and popular protest},
  author={{Economist Intelligence Unit}},
  year={2020},
  url = {https://www.eiu.com/topic/democracy-index}
}

@article{listCountriesCapitals,
    title = {List of Countries and Capitals},
    author = {{Techslides.com}},
    year = {2016},
    url = {http://techslides.com/list-of-countries-and-capitals}
}

@article{bass1969new,
  title={A new product growth for model consumer durables},
  author={Bass, Frank M},
  journal={Management science},
  volume={15},
  number={5},
  pages={215--227},
  year={1969},
  publisher={INFORMS}
}

@article{scheffer2009early,
  title={Early-warning signals for critical transitions},
  author={Scheffer, Marten and Bascompte, Jordi and Brock, William A and Brovkin, Victor and Carpenter, Stephen R and Dakos, Vasilis and Held, Hermann and Van Nes, Egbert H and Rietkerk, Max and Sugihara, George},
  journal={Nature},
  volume={461},
  number={7260},
  pages={53--59},
  year={2009},
  publisher={Nature Publishing Group}
}

@article{kvasnivcka2014viral,
  title={Viral video diffusion in a fixed social network: an agent-based model},
  author={Kvasni{\v{c}}ka, Michal},
  journal={Procedia Economics and Finance},
  volume={12},
  pages={334--342},
  year={2014},
  publisher={Elsevier}
}

@article{chen2019agent,
  title={An agent-based model for information diffusion over online social networks},
  author={Chen, Zhuo},
  journal={Papers in Applied Geography},
  volume={5},
  number={1-2},
  pages={77--97},
  year={2019},
  publisher={Taylor \& Francis}
}

@article{fuchs2022covid,
  title={Covid-Induced School Closures in the US and Germany: Long-Term Distributional Effects},
  author={Fuchs-Sch{\"u}ndeln, Nicola},
  year={2022},
  publisher={CEPR Discussion Paper No. DP17205}
}

\end{document}